\documentclass[twocolumn]{aastex63}
\submitjournal{BAAS}
\usepackage{CJKutf8}

\usepackage[hyphens]{url}
\usepackage{hyperref}
\usepackage[hyphenbreaks]{breakurl}

\shorttitle{A Referee Primer}
\shortauthors{Ntampaka et al.}

\begin{document}

\correspondingauthor{Michelle Ntampaka}
\email{mntampaka@stsci.edu}

\author[0000-0002-0144-387X]{Michelle Ntampaka}
\affiliation{Space Telescope Science Institute, Baltimore, MD 21218,USA}
\affiliation{Department of Physics \& Astronomy, Johns Hopkins University, Baltimore, MD 21218,USA}

\author[0000-0002-7846-9787]{Ana Bonaca}
\affiliation{The Observatories of the Carnegie Institution for Science, 813 Santa Barbara St., Pasadena, CA 91101, USA}

\author[0000-0002-0974-5266]{Sownak Bose}
\affiliation{Institute for Computational Cosmology, Department of Physics, Durham University, Durham DH1 3LE, UK}

\author[0000-0002-2929-3121]{Daniel J.\ Eisenstein}
\affiliation{Center for Astrophysics $|$ Harvard \& Smithsonian, 60 Garden St., Cambridge, MA 02138, USA}

\author[0000-0002-2929-3121]{Boryana Hadzhiyska}
\affiliation{Center for Astrophysics $|$ Harvard \& Smithsonian, 60 Garden St., Cambridge, MA 02138, USA}

\author[0000-0002-3407-1785]{Charlotte Mason}
\affiliation{Cosmic Dawn Center (DAWN), Denmark}
\affiliation{Niels Bohr Institute, University of Copenhagen, Jagtvej 128, DK-2200 Copenhagen N, Denmark}

\author[0000-0002-6766-5942]{Daisuke Nagai}
\affiliation{Department of Physics, Yale University, New Haven, CT 06520, USA}
\affiliation{Department of Astronomy, Yale University, New Haven, CT 06520, USA}

\author[0000-0003-2573-9832]{Joshua S. Speagle (\begin{CJK*}{UTF8}{gbsn}沈佳士\ignorespacesafterend\end{CJK*})}
\altaffiliation{Banting \& Dunlap Fellow}
\affiliation{Department of Statistical Sciences, University of Toronto, Toronto, ON M5S 3G3, Canada}
\affiliation{David A. Dunlap Department of Astronomy \& Astrophysics, University of Toronto, Toronto, ON M5S 3H4, Canada}
\affiliation{Dunlap Institute for Astronomy \& Astrophysics, University of Toronto, Toronto, ON M5S 3H4, Canada}

\title{A Referee Primer for Early Career Astronomers} 

\begin{abstract}
    Refereeing is a crucial component of publishing astronomical research, but few professional astronomers receive formal training on how to effectively referee a manuscript.  In this article, we lay out considerations and best practices for referees.  This document is intended as a tool for early career researchers to develop a fair, effective, and efficient approach to refereeing.
 \\
    
\end{abstract}

\section{Introduction}

Referees are responsible for assessing the quality and novelty of research and to provide feedback to the editor on whether the result is appropriate for the journal.  They are also responsible to provide feedback on the framing and presentation of the results.  An effective referee will respond with a fair, kind, and actionable report to the authors and to the journal editor.

Because they provide support and feedback for assessing new scientific results, referees are a vital part of scientific publication. And though this is an important part of our scientific careers, few astronomers receive formal training in the process.  

In this document, we discuss the process of refereeing, including: understanding ethical conflicts, framing the referee report, avoiding common referee pitfalls, navigating rejections, and framing the final referee report.  See \cite{survive-peer,  nicholas2011quick, Raff-ref}, and references therein, for additional information and complementary perspectives on refereeing best practices.

This reference is intended primarily for referees-in-training as they learn how to referee journal articles.  It may also be useful to mentors as a tool for coaching their early-career mentees through the process of becoming an effective referee.   

\section{Understanding Ethical Conflicts}
 
Before agreeing to referee a manuscript, consider ethical conflicts that might prevent a referee from being a fair referee or that might damage your professional reputation and relationships.

\begin{enumerate}
\item  Am I a close collaborator of any of the authors?  Engaging in an anonymous, critical conversation with close collaborators may damage these professional relationships.
\item  Do I have a poor relationship with any of the authors?  Even if you feel that you can give an unbiased assessment of their work, remember that there are other valid referees who would not be spending a large amount of emotional energy to fairly assess the work.
\item  Is it ``too close to comfort''?   If the research is so close to your own work in progress that you could not give honest feedback or suggestions for improvement, you should discuss this with the editor before proceeding.
\item  Do I have the expertise to properly assess this research?  This can be particularly difficult in the case of work that draws from more than one discipline, and any concerns about  interdisciplinary work should be directed to the editor before you agree to review the manuscript.
\end{enumerate}
If you suspect that you cannot be a fair, objective, and fully invested referee for any reason, you should write a short note to the editor to discuss your concerns; the editor can help you to evaluate this.
 
\section{Questions for Consideration}
Expectations and norms vary from journal to journal \citep[e.g.][]{AAS-ref, MNRAS-ref, Nature-ref}, and when possible, you should consult with the editorial staff about what qualities should be considered in your review.  Assessment questions we have found useful to focus on are:
\begin{enumerate}
\item  Is the research sufficiently innovative?  Does it bring something new to the field?
\item  Do the authors put the research in sufficient context?  Is their literature review sufficient?  Have they missed any citations? 
\item  Are the data and methods clearly described?  Is there enough information to recreate the authors’ analysis?
\item  Have the authors provided sufficient information to make their results reproducible, including access to data and software?  The referee should assess whether the authors have an adequate plan for sharing data and software with the community and should share this assessment with the editor.
\item  Are the methods used appropriately? 
\item  Are there passages that are unclear or ambiguous?
\item  Have the authors provided sufficient evidence to support their claims?  Have the authors made unsubstantiated claims?  Have they overstated their results?
\item  Have the authors appropriately discussed caveats and limitations?  Have they provided clear explanations for any results that seem too good to be true?
\end{enumerate}
While it is not necessary to answer each of these questions explicitly, these are just some of the assessments that will frame your report.

\section{Avoiding Common Pitfalls} 
As a referee, you should not make unkind criticisms about topics outside of the research and manuscript you are assessing.  Do not make statements about the authors or their qualifications.

You should not serve as a copy editor.  If significant language issues make it difficult to assess the scientific content of the manuscript, pause your review.  Recommending ``proofreading by a native speaker'' is inappropriate.   Instead, you should immediately contact the editor, tell them your concerns, and let them reach out to the authors.  If the main body of the manuscript is sufficiently understandable but there are typos or ambiguous passages, commenting on these is fine.

As a referee, you should not expand the scope of the manuscript or push the manuscript to deviate significantly from the authors' intentions.  You are not the authors' academic advisor and it is inappropriate to change the direction or significantly expand the scope of the manuscript.  There are two exceptions to this:  1. if the original scope is insufficient for publication, you can and should explain this and consider recommending ways to extend the work, and  2.  If you see a golden opportunity for the authors to improve their work,  you might offer this suggestion but be clear that this is not a requirement for publication.

You should not write meandering prose regarding your opinions on the topic.  The authors will need to respond to each item in your report, and this report is most useful when it is focused on clear, actionable items.
 
\section{Navigating Rejections} 
It can be more difficult to recommend rejection than to recommend revision and resubmission.  In the case of rejections, you will need to write a clear and kind note to the authors explaining the shortfalls of the manuscript, including strong evidence supporting your claim that it is not appropriate for the journal.  This does not need to be exhaustive, but you do need to be specific about some critical flaws.  It can be difficult to write this in a way that is both kind and constructive, but this should be your goal.  Feedback that is not constructive can be included in the confidential response to the editor.

\section{Framing the Final Report} 
Begin your review by explicitly describing the strengths, aims, and results of the manuscript.  This sets a good tone for the constructive criticism that follows and it communicates to the authors that you understand their research goals.  By including a summary of the manuscript and listing its strengths (rather than focusing only on the manuscript's shortcomings), you  communicate fairness.  Using ``the manuscript'' instead of ``the author'' or ``you'' in your feedback is one way to keep your feedback neutral.

Organize your feedback.  The following template is commonly used: 
\begin{enumerate}
\item  State the aims and key results of manuscript.
\item  Summarize the strengths of the manuscript.
\item  Summarize your constructive feedback or assessment of the manuscript.
\item  State your recommendation for publication.
\item  List the major weaknesses (e.g., methodological issues or overstated results).
\item  List the minor weaknesses (e.g., missing references or figure formatting issues).
\end{enumerate}

Clearly articulate your feedback.  Refer to sections and line numbers where applicable, and provide sufficient publication information (or arxiv or ADS links) for the authors to quickly find references.  

\section{Final Comments} 

Before you submit your report, you should edit the first draft of your report for tone.  This is particularly important because emails often come across as more harsh than they are intended.  Aim for polite discussion with constructive criticism and clear, actionable items.   Your report should be courteous without compromising honesty.

Referees provide an important service to our scientific community: evaluating new research and providing feedback to ensure that the research is clearly presented and appropriately framed.  As a referee, your goal should be to provide the type of report that you would want to receive.  Strive to write a report that is a fair, kind, and actionable assessment of the research.  
\bigskip
\bigskip
\bigskip

\acknowledgements{}
We thank the anonymous member of the AAS Publications Editorial Staff for reviewing this article and for providing valuable feedback. We also thank N\'estor \mbox{Espinoza}, Susan Mullally, and Laura Watkins for providing thoughtful feedback on this document.

\bibliography{references}

\begin{thebibliography}{}
\expandafter\ifx\csname natexlab\endcsname\relax\def\natexlab#1{#1}\fi
\providecommand{\url}[1]{\href{#1}{#1}}
\providecommand{\dodoi}[1]{doi:~\href{http://doi.org/#1}{\nolinkurl{#1}}}
\providecommand{\doeprint}[1]{\href{http://ascl.net/#1}{\nolinkurl{http://ascl.net/#1}}}
\providecommand{\doarXiv}[1]{\href{https://arxiv.org/abs/#1}{\nolinkurl{https://arxiv.org/abs/#1}}}

\bibitem[{{American Astronomical Society}(2022)}]{AAS-ref}
{American Astronomical Society}. 2022, Information for Referees.
\newblock
  \url{iopscience.iop.org/journal/1538-3881/page/Information\%20for\%20referees}

\bibitem[{{MNRAS}(2022)}]{MNRAS-ref}
{MNRAS}. 2022, Instructions to Authors.
\newblock \url{https://academic.oup.com/mnras/pages/General_Instructions}

\bibitem[{{Nature}(2022)}]{Nature-ref}
{Nature}. 2022, Guide to Referees.
\newblock \url{https://www.nature.com/srep/guide-to-referees}

\bibitem[{Nicholas \& Gordon(2011)}]{nicholas2011quick}
Nicholas, K.~A., \& Gordon, W.~S. 2011, Eos, Transactions American Geophysical
  Union, 92, 233

\bibitem[{{Raff}(2013)}]{Raff-ref}
{Raff}, J. 2013, How to become good at peer review: A guide for young
  scientists.
\newblock
  \url{https://violentmetaphors.com/2013/12/13/how-to-become-good-at-peer-review-a-guide-for-young-scientists/}

\bibitem[{{Wager} {et~al.}(2002){Wager}, {Godlee}, \&
  {Jefferson}}]{survive-peer}
{Wager}, E., {Godlee}, F., \& {Jefferson}, T. 2002, How to Survive How to
  Survive Peer Review (BMJ Books)

\end{thebibliography}

\end{document}